\magnification 1200
\baselineskip = 15pt
\hsize=5.7in
\hoffset=-10pt

\font\BLarge=cmbx12
\font\Frak=eufm10
\def\frak#1{{\hbox{\Frak#1}}}

\font\Bbb=msbm10 
\def\BBB#1{\hbox{\Bbb#1}}

\font\petit=cmr10 scaled 800

\font\funny=cmr12 scaled\magstep 3
\def\leftno{{\kern 0.2em {\funny :} \kern-0.08em}}
\def\rightno{{\kern-0.08em {\funny :} \kern 0.2em}}

\hyphenation{Pet-via-shvili}

\def\g{{\frak g}}
\def\h{{\frak h}}
\def\s{{\frak s}}
\def\C{\BBB C}
\def\Z{\BBB Z}
\def\N{\BBB N}
\def\R{\BBB R}
\def\odd{{\BBB N}_{\rm odd}}

\def\dg{\dot \g}
\def\ds{\dot \s}

\def\t{{\bf t}}
\def\r{{\bf r}}
\def\m{{\bf m}}
\def\q{{\bf q}}
\def\x{{\bf x}}
\def\u{{\bf u}}
\def\v{{\bf v}}
\def\n{{\bf n}}
\def\y{{\bf y}}
\def\A{{\cal A}}
\def\e{\epsilon}

\def\tx{{\tilde x}}
\def\tu{{\tilde u}}
\def\tv{{\tilde v}}

\def\tw{{\tilde w}}
\def\ty{{\tilde y}}
\def\btx{{\bf \tilde x}}
\def\bty{{\bf \tilde y}}
\def\btu{{\bf \tilde u}}
\def\btv{{\bf \tilde v}}
\def\exp#1{{\rm exp} \left( #1 \right)}

\def\hgt{{\rm ht}}
\def\mod{\hbox{\rm mod \ }}
\def\ol{\overline}
\def\phi{\varphi}
\def\d{\partial}

\def\Der{{\rm Der}}
\def\K{{\cal K}}
\def\Ct{\C[t_1^\pm,\ldots, t_n^\pm]}
\def\Diff{{\it Diff}}
\def\p{\prime}
\def\pp{{\prime\prime}}
\def\semid{\hbox{\vrule width 0.5pt height 5pt 
\kern -1.3pt $\times$}}

\

\

\

\centerline{\BLarge An Extension of the KdV Hierarchy} 
\centerline{\BLarge Arising from a Representation of a 
Toroidal Lie Algebra}
\footnote{}{1991 {\it Mathematics Subject Classification.}
Primary 17B69, 35Q51.} 

\

\centerline{{\bf Yuly Billig}
\footnote{*}{Work supported by the Natural Sciences and 
Engineering Research Council of Canada.}}

\

\centerline{Department of Mathematics and Statistics,
University of New Brunswick,}
\centerline{Fredericton, N.B., E3B 5A3, Canada} 

\

\vskip 1cm

{
\leftskip=0.5in
\rightskip=0.5in
\petit
\noindent
ABSTRACT. In this article we show how to construct
hierarchies of partial differential equations from
the vertex operator representations of toroidal
Lie algebras. In the smallest example - rank 2 
toroidal cover of $sl_2$ - we obtain an
extension of the KdV hierarchy. We use the action
of the corresponding infinite-dimensional group
to construct solutions for these non-linear PDEs. 
\par}

\vskip 1cm

\

{\bf 0. Introduction.}

 In this article we show how to construct hierarchies of
partial differential equations and their soliton-type
solutions from the vertex operator representations
of toroidal Lie algebras.

Soliton theory was given a new impetus when it was 
linked with the representation theory of 
infinite-dimensional Lie algebras in the works of
 Sato [S], Date-Jimbo-Kashiwara-Miwa [DJKM] and 
Drinfeld-Sokolov [DS]. It was discovered
that for various partial differential equations 
the space of soliton solutions has a large group
of hidden symmetries. Moreover, for every Kac-Moody
algebra one can construct a hierarchy of PDEs whose
symmetries form the corresponding Kac-Moody group [KW].

The most famous example occurs in the context of the
affine Kac-Moody algebra $A_1^{(1)}$ which is a 
central extension of the loop algebra 
$sl_2(\C[t,t^{-1}])$. In this case the hierarchy 
contains the Korteweg-de Vries equation
$$ f_t = f f_x + f_{xxx}$$
as the equation of the lowest degree.

We generalize this approach for the toroidal 
Lie algebras. The difficulties arise because
the toroidal Lie algebras have no
triangular decomposition and many methods of the
Kac-Moody theory do not work. However, an important 
class of representations of these algebras was
constructed in [MEY], [EM] and [B]. Here we establish 
a connection between  the principal vertex operator 
realization developed in [B] and non-linear partial 
differential equations.

We study in detail the case of the smallest toroidal
algebra - the universal central extension of
$sl_2(\C[t_0,t^{-1}_0,t_1,t^{-1}_1])$. This algebra
has affine Kac-Moody algebra $A_1^{(1)}$ as a 
subalgebra. The hierarchy we obtain has the KdV
hierarchy as a proper subset. Equations of low
degrees in the extended KdV hierarchy, but not in
the KdV subhierarchy are the following:
$$ {\d\over\d x} \left( f_t - 
{1\over 6} f_{xxy} - f_x f_y \right) = 
{1\over 6} f_{yz} $$
and
$$ f_{xxxt} + 6 f_{xx} f_{xt} - f_{xxyz} - 4 f_{xy} f_{xz} - 2 f_{xx} f_{yz} = 0 .$$
We use algebraic methods
to construct solutions for the extended KdV hierarchy. 

% The software package MapleV is very efficient for
% the calculation of the hierarchy of PDEs.

The paper is organized as follows. In Section 1 we 
present the construction of the toroidal algebras.
In Section 2 we review the main results of [B] on the
principal vertex operator representations. We obtain 
the extended KdV hierarchy in Section 3 and construct
its solutions in Section 4. In the Appendix we discuss
the generalized Casimir operators and give a proof
of Proposition 1 which is fairly standard
but crucial for our derivation.

\

{\bf 1. Toroidal Lie algebras.}

Throughout this paper we will use the constructions and 
the notations of [B]. 

Let $\dg$ be a finite-dimensional simple Lie algebra
over $\C$ of type $A_\ell$, $D_\ell$  or $E_\ell$
 (i.e., simply-laced) with the root system $\dot\Delta$.
The algebra $\dg$ possesses
a non-degenerate symmetric invariant bilinear form 
$( \cdot | \cdot )$ . The reduction of this form to 
the Cartan subalgebra $\dot\h$ induces the map $\nu : \dot\h \rightarrow \dot\h^*$ and a bilinear form on 
$\dot\h^*$. 
We normalize both forms by the condition
$(\alpha | \alpha)= 2$ for all nonzero roots $\alpha \in 
\dot \Delta$. Let $\{ \alpha_1,\ldots,\alpha_\ell \}$
be the simple roots in $\dot \Delta$. Define the height
function $\hgt : \ \dot\Delta \rightarrow \Z$ by 
$\hgt (\sum\limits_{j=1}^\ell k_j \alpha_j ) = 
\sum\limits_{j=1}^\ell k_j $. Let $\rho\in\dot\h^*$ be
such that $(\rho | \alpha ) = \hgt(\alpha)$ for 
$\alpha\in\dot\Delta$.  

Let $h$ be the Coxeter number of $\dg$. Consider the
principal $\Z_h$-gradation of $\dg$
$$\dg = \sum\limits_{j\in\Z_h} \dg_j ,$$
where $\dg_j$ is the direct sum of the root spaces
$\dg^\alpha$ with $\hgt(\alpha) = j (\mod h)$.

The algebra $\dg$ possesses a Cartan subalgebra
$\ds$ which is homogeneous with respect to the principal
$\Z_h$-gradation and has a basis $\{ T_1,\ldots,T_\ell\}$
such that $T_i \in \dg_{m_i}$ , where the numbers
$1=m_1\leq m_2 \leq \ldots \leq m_\ell = h-1$ are
the exponents of $\dg$. Define a sequence 
$\{ b_i \}_{i\in\N}$ by $b_{j\ell + i} = jh + m_i$
for $i=1,\ldots,\ell$ and $j\geq 0$. 

The basis of $\dot\s$ can be normalized so that
$$ (T_i | T_{\ell+1-j} ) = h\delta_{ij} . \eqno{(1.1)}$$

Let $\dot\Delta_s$ be the root system of $\dg$ with 
respect to the Cartan subalgebra $\ds$. For $\alpha\in
\dot\Delta_s$ fix a root element 
$$A^\alpha = \sum\limits_{j\in\Z_h} A^\alpha_j ,
\hbox{\hskip 0.7cm} A^\alpha_j \in \dg_j.$$
Define constants $\lambda^\alpha_i = \alpha(T_i)$.
Then $[T_i, A^\alpha] = \lambda^\alpha_i A^\alpha$.

% Let $\zeta\in\C$ be a primitive $h$-th root of 1.
% Note that for a root element $A^\alpha$,
% $$ \sum\limits_{j\in\Z_h} \zeta^j A^\alpha_j$$
% is also a root element. This gives a free $\Z_h$-action 
% on $\dot\Delta_s$. Fix a set $\{\beta_1,\ldots,
% \beta_\ell \}$ of the representatives in these orbits.

% $$\dg_j =  \sum\limits_{\alpha\in\dot\Delta \atop
% \hgt(\alpha) = j (\mod h)} \dg^\alpha .$$

 A central extension of the Lie algebra
$$\sum\limits_{j\in\Z} \dg_j \otimes s^j $$
is the principal realization of untwisted affine Kac-Moody
algebra. The $n+1$-toroidal Lie algebra is the universal
central extension of
$$\sum\limits_{j\in\Z} \dg_j \otimes s^j 
\Ct.$$
The vertex operator representations of this algebra
were studied in [MEY], [EM] and [B]. In the present paper
we will use a larger algebra
$$\tilde\g = \sum\limits_{j\in\Z} \dg_j \otimes 
s^j \C[\R^n] .$$
We replace the algebra of Laurent polynomials $\Ct$
which is the group algebra of $\Z^n$
with the group algebra of $\R^n$. The algebra $\C[\R^n]$
has a basis of monomials $\t^\r = t_1^{r_1} \ldots 
t_n^{r_n} , \r = (r_1,\ldots, r_n) \in \R^n$ . 
The multiplication in $\C[\R^n]$ is given by $\t^\r
\t^\m = \t^{\r+\m}$. All the results (and their proofs)
from [B] remain valid for this version of the toroidal
Lie algebras.

The following description of the universal central 
extension of  $\tilde \g$ is based on the general result 
of [Kas] (see also [MEY]).

Let $\dot \K$ be an $(n+1)$-dimensional space with the 
basis $\{ K_0, K_1,\ldots,K_n \}$. Consider the space
$$\tilde \K = \dot\K \otimes \C[s^h,s^{-h}]\otimes
\C[\R^n] $$
and its subspace $d\tilde\K$ spanned by the elements
$$r_0 K_0 \otimes s^{r_0 h} \t^\r + 
r_1 K_1  \otimes s^{r_0 h} \t^\r + \ldots +
r_n K_n  \otimes s^{r_0 h} \t^\r .$$
The factor space $\K = \tilde\K / d\tilde\K$ is the space
of the universal central extension of $\tilde\g$ and the
Lie bracket in the toroidal algebra $\g = \tilde\g 
\oplus \K$ is given by 
$$ [g_1 \otimes f_1 (s,\t), g_2 \otimes f_2 (s,\t)] =
[g_1, g_2] \otimes (f_1 f_2) + (g_1 | g_2)
\left\{ {1\over h} s{\d f_1 \over \d s} f_2 K_0 +
\sum\limits_{p=1}^n t_p {\d f_1 \over \d t_p} f_2 K_p 
\right\}$$
and 
$$ [\g , \K] = 0 .$$
%
% We denote by $s^{r_0 h} \t^\r K_p$ the image of
% $K_0 \otimes s^{r_0 h} \t^\r$ in $\K$. 
%
From now on we
will omit the tensor product sign when writing the
elements of $\g$.

\

{\bf 2. Principal vertex operator construction for
toroidal Lie algebras.}

Now we can describe the representation of $\g$ constructed
in [B]. The space $F$ of this representation (the Fock
space) is the tensor product of the group algebra 
of $\R^n$ (with the basis $\{ \q^\r | \r\in\R^n \}$ )
and the algebra of polynomials in infinitely many 
variables:
$$ F = \C[\R^n] \otimes \C[x_{b_i}, u_{pi}, 
v_{pi} ]^{p=1,\ldots,n}_{i\in\N} .$$
Instead of specifying the action of individual elements
of $\g$ on $F$, we will represent  certain generating 
series (by vertex operators). Let $z$ be a formal 
variable. We set
$$ \sum\limits_{j\in\Z} \phi (s^{jh} \t^\r K_0) z^{-jh} = 
K_0 (z, \r) , \hbox{ \ \ \ where} \eqno{(2.1)}$$
$$K_0 (z, \r) = \q^\r \exp{ \sum\limits_{p=1}^n r_p 
\sum\limits_{j\geq 1} z^{jh} u_{pj} }
\exp{ - \sum\limits_{p=1}^n r_p 
\sum\limits_{j\geq 1} {z^{-jh}\over j} 
{\d \over \d v_{pj}} } ,\eqno{(2.2)}$$
$$\sum\limits_{j\in\Z} \phi (s^{jh} \t^\r K_p) z^{-jh} = 
K_p(z, \r) =
K_p(z) K_0 (z, \r) , \hbox{ \ \ \ where}\eqno{(2.3)}$$
$$K_p(z) = \sum\limits_{i\geq 1} i z^{ih} u_{pi} 
+ \sum\limits_{i\geq 1} z^{-ih} {\d \over \d v_{pi}} ,
\eqno{(2.4)}$$
$$\sum\limits_{j\in\Z} \phi (T_i s^{m_i + jh} \t^\r)
z^{-m_i - jh} = T_i (z, \r) =
T_i (z) K_0(z,\r), \hbox{ \ \ } 
i = 1, \ldots , \ell , \hbox{ \ \ \ where}
\eqno{(2.5)}$$
$$ T_i(z) = \sum\limits_{j\geq 1} (jh-m_i) z^{jh-m_i}
 x_{jh-m_i} + \sum\limits_{j\geq 0} z^{-jh-m_i}
{\d \over \d x_{jh + m_i}} , \eqno{(2.6)}$$
$$\sum\limits_{j\in\Z} \phi ( A^\alpha_j s^j \t^\r ) 
z^{-j} = A^\alpha (z, \r) =
A^\alpha (z) K_0 (z, \r) , \hbox{ \ \ }
\alpha\in\dot\Delta_s , \hbox{ \ \ \ where}
\eqno{(2.7)}$$
$$A^\alpha(z) = - {\rho(A^\alpha_0) \over h}
\exp{ \sum\limits_{i\geq 1} \lambda^\alpha_i z^{b_i} 
x_{b_i} } 
\exp{ - \sum\limits_{i\geq 1} \lambda^\alpha_{\ell+1-i} 
{z^{-b_i}\over b_i} {\d \over \d x_{b_i} }} .
\eqno{(2.8)}$$

We also represent derivations of $\g$ as operators on $F$.
Before we introduce these, we need to discuss the 
operation of the normal ordering.

Consider the algebra of the differential operators 
$\Diff(y_1, y_2,\ldots)$ on the 
space $\C[y_1,y_2,\ldots]$:
$$ \Diff(y_1, y_2,\ldots) = \left\{
\sum\limits_{\n\in\A} f_\n (\y) \left( {\d\over\d \y}
\right)^\n \big| f_\n (\y)\in\C[y_1,y_2,\ldots] \right\} ,
$$
where $\A$ is the set of sequences
$\n = (n_1, n_2, \ldots )$ with $n_i \in\Z_+$ where
only finitely many terms are nonzero. We use the
notations $ \y = ( y_1, y_2, \ldots )$ and
$\left( {\d\over\d \y} \right)^\n =
\left( {\d\over\d y_1} \right)^{n_1}
\left( {\d\over\d y_2} \right)^{n_2} \ldots$.
A differential operator $P\in\Diff(y_1, y_2,\ldots)$
can be viewed as a linear map
$$ P : \hbox{\hskip 0.5cm} 
\C[y_1,y_2,\ldots] \rightarrow \C[y_1,y_2,\ldots] .$$
The space $\Diff(y_1, y_2,\ldots)$ has a structure
of an associative algebra with respect to the composition
of operators.  This algebra is
not commutative since ${\d \over \d y_i}$
and $y_i$ do not commute. The normal ordering \ : \ : \ is a 
new commutative associative product on
$\Diff(y_1, y_2,\ldots)$ defined by
$$\leftno \left( \sum\limits_{\n\in\A} f_\n (\y) \left( 
{\d\over\d \y} \right)^\n \right)
\left( \sum\limits_{\n\in\A} g_\m (\y) \left( 
{\d\over\d \y} \right)^\m \right) \rightno
= \sum\limits_{\n+\m\in\A} f_\n (\y) g_\m (\y) \left( 
{\d\over\d \y} \right)^{\n+\m} .$$

Now consider the following operators on $F$:
$$D_p (z, \r) = \leftno D_p(z) K_0(z,\r) \rightno ,
 \hbox{\ \ } p = 1, \ldots , n, \hbox{\ , \ \ where}
\eqno{(2.9)}$$
$$D_p(z) = \sum\limits_{i\geq 1} i z^{ih} v_{pi} +
q_p{\d\over \d q_p} + 
\sum\limits_{i\geq 1} z^{-ih} {\d\over\d u_{pi}}, 
\hbox{ \ \ and} \eqno{(2.10)}$$
$$D_s (z, \r) = \leftno D_s(z) K_0(z,\r)\rightno \hbox{\ , \ \ where}
\eqno{(2.11)}$$
$$D_s(z) = -{1\over 2} \sum\limits_{i=1}^\ell
\leftno T_i(z) T_{\ell+1-i}(z)\rightno - h\sum\limits_{p=1}^n
\leftno D_p(z) K_p(z) \rightno . \eqno{(2.12)}$$
 Expanding the generating series $D_p(z,\r)$ and
$D_s(z,\r)$, we obtain operators $s^{jh}\t^\r D_p$ and
\break $s^{jh}\t^\r D_s$:
$$D_p(z,\r) = \sum\limits_{j\in\Z} s^{jh}\t^\r D_p z^{-jh}
\hbox{\ \ , \ \ } 
D_s(z,\r) = \sum\limits_{j\in\Z} s^{jh}\t^\r D_s
z^{-jh}. \eqno{(2.13)}$$

 We summarize below the main results (Theorem 5 and 
Proposition 8) of [B].

{\bf Theorem A.}
(a). 
{\it
The formulas (2.1)-(2.8) define a representation 
of the toroidal Lie algebra $\g$ on the Fock space $F$.
}

\noindent
(b). 
{\it
The operators $s^{r_0 h}\t^\r D_p$ and 
$s^{r_0 h}\t^\r D_s$
defined by (2.9)-(2.13) act on $\g$ as derivations:
}
$$\left[ s^{r_0 h}\t^\r D_p, \phi\left( A^\alpha_j
s^j \t^\m \right) \right] = 
m_p \phi\left( A^\alpha_j s^{r_0h + j} \t^{\r+\m} \right)
 ,$$
$$\left[ s^{r_0 h}\t^\r D_s, \phi\left( A^\alpha_j
s^j \t^\m \right) \right] = 
j \phi\left( A^\alpha_j s^{r_0h + j} \t^{\r+\m} \right)
 ,$$
$$\left[ s^{r_0 h}\t^\r D_p, \phi\left( T_i
s^{b_i} \t^\m \right) \right] = 
m_p \phi\left( T_i s^{r_0h + b_i} \t^{\r+\m} \right)
 ,$$
$$\left[ s^{r_0 h}\t^\r D_s, \phi\left( T_i
s^{b_i} \t^\m \right) \right] = 
b_i \phi\left( T_i s^{r_0h + b_i} \t^{\r+\m} \right) .$$

% 
% $$\left[ s^{r_0 h}\t^\r D_p, \phi\left( 
% s^{m_0 h} \t^\m  K_a \right) \right] = \phi\left(
% m_p s^{(r_0 + m_0)h} \t^{\r+\m} K_a
% + \delta_{ap} \sum\limits_{b=0}^n r_b 
% s^{(r_0 + m_0)h} \t^{\r+\m} K_b
% \right), $$
% $$\left[ s^{r_0 h}\t^\r D_s, \phi\left( 
% s^{m_0 h} \t^\m  K_a \right) \right] = h \phi\left(
% m_0 s^{(r_0 + m_0)h} \t^{\r+\m} K_a
% + \delta_{a0} \sum\limits_{b=0}^n r_b 
% s^{(r_0 + m_0)h} \t^{\r+\m} K_b \right)
% . $$
% 

{\bf Remark.} Though the operators 
$s^{r_0 h}\t^\r D_p$ and $s^{r_0 h}\t^\r D_s$ 
act on $\phi(\g)$
as derivations of $\C[s^h, s^{-h}] \otimes \C [\R^n]$,
the algebra they generate
together with $\phi(\g)$ is not isomorphic to the
semidirect product of $\g$ with ${\cal D} = \Der 
\left( \C[s^h, s^{-h}] \otimes \C [\R^n] \right)$ . 
A direct computation shows that 
the span of $s^{r_0 h}\t^\r D_p$ and $s^{r_0 h}\t^\r 
D_s$ is not closed under the Lie bracket. The commutators
contain extra terms that commute with $\phi(\g)$.

Our main tool for the construction of the hierarchies of partial differential equations will be the generalized Casimir operators. We introduce these by the following generating series (see Appendix for details):
$$\Omega(z) = \sum\limits_{j\in\Z} \Omega_j z^{-jh} =$$
$$ = \leftno \bigg\{ {1\over h} 
\sum\limits_{i=1}^\ell T_i(z)\otimes T_{\ell+1-i}(z)
+ \sum\limits_{\alpha\in\dot\Delta_s}
 { 1\over ( A^\alpha | A^{-\alpha} ) }  
A^\alpha (z) \otimes A^{-\alpha} (z) 
- {\ell ( h + 1 ) \over 12 h}$$
$$ + {1\over h} D_s (z) \otimes 1 + {1\over h} 1 \otimes
D_s (z) + \sum\limits_{p=1}^n D_p (z) \otimes K_p (z)
+  \sum\limits_{p=1}^n K_p (z) \otimes D_p (z) 
 \bigg\}
\times $$
$$ \times \sum\limits_{\r\in\R} K_0(z,\r) \otimes 
K_0(z,-\r) \rightno .\eqno{(2.14)}$$

 Both $\g$ and its module $F$ are graded by $\Z \times
\R^n$. Thus the $U(\g)\otimes U(\g)$-module 
$F\otimes F$ is graded by $(\Z \times \R^n) \times
(\Z \times \R^n)$ and we can consider its completion
$\overline{F\otimes F}$ with respect to this grading.
The operators $\Omega_k$ act from $F\otimes F$ to
$\overline{F\otimes F}$. Also note that both
$F\otimes F$ and $\overline{F\otimes F}$ have the 
$\g$-module structure.

{\bf Proposition 1.} 
{\it 
The operators 
$$\Omega_k : \hbox{\hskip 0.5cm}
 F\otimes F \rightarrow \overline{F\otimes F}$$
commute with the action of $\g$.
}

The proof of this proposition is given in the Appendix.

\

{\bf 3. Extended KdV hierarchy.}

 As in the affine case, the equation $\Omega_k (\tau
\otimes \tau ) = 0$ decomposes in the hierarchy
of partial differential equations in the Hirota
form. In this section we study the hierarchy that 
corresponds to the smallest toroidal algebra with
$\dg = sl_2(\C)$ and $n=1$. The Coxeter number of
$sl_2(\C)$ is $h = 2$ and its only exponent is
$m_1 = 1$. 

 The representation space is
$$ F = \C[\R] \otimes \C[x_1,x_3,x_5,\ldots] \otimes
\C[u_1, u_2,\ldots] \otimes \C[v_1,v_2,\ldots] ,$$
where $\C[\R]$ is the group algebra of $(\R, +)$ with
the basis $\{ q^r | r\in\R \}$. The action (2.1)-(2.8) 
of $\g$ on $F$ can be written as follows (cf. [Kac], 
Sec. 14.13):
$$ K_0(z,r) = q^r \exp{r\sum\limits_{i\in\N} z^{2i} u_i}
\exp{ -r\sum\limits_{i\in\N} {z^{-2i}\over i} 
{\d\over\d v_i} },$$
$$ K_1 (z) = \sum\limits_{i\in\N} iz^{2i} u_i + 
\sum\limits_{i\in\N} z^{-2i} {\d\over\d v_i} ,$$
$$ T(z) = \sum\limits_{j\in\odd} j z^j x_j +
\sum\limits_{j\in\odd} z^{-j} {\d\over\d x_j} ,$$
$$ A^{\pm\alpha} (z) = {1\over 2}
\exp{ \pm 2\sum\limits_{j\in\odd} z^j x_j }
\exp{ \mp 2\sum\limits_{j\in\odd} {z^{-j}\over j} 
{\d\over\d x_j }} ,$$
$$ T(z,r) = T(z) K_0(z,r) , \hbox{\ \ }
 A^{\pm\alpha}(z,r) = A^{\pm\alpha}(z) K_0(z,r) , 
\hbox{\ \ } K_1(z,r) = K_1(z) K_0(z,r) .$$
For the positive root $\alpha$ of $sl_2(\C)$,
we will denote $A^\alpha (z, r)$
simply by $A (z,r)$. Then $A^{-\alpha} (z,r) = A(-z,r)$.

The derivations of $\g$ are represented by
$$ D_1 (z) = \sum\limits_{i\in\N} iz^{2i} v_i + 
q{\d\over\d q} +
\sum\limits_{i\in\N} z^{-2i} {\d\over\d u_i} ,$$
$$ D_s (z) = -{1\over 2} \leftno T(z) T(z)\rightno - 
2 \leftno D_1(z) K_1(z)\rightno ,$$
$$ D_1 (z,r) = \leftno D_1(z) K_0(z,r)\rightno ,  
\hbox{\hskip 0.5cm}
 D_s (z,r) = \leftno D_s(z) K_0(z,r)\rightno .$$

The Casimir generating series (2.14) can be written as
$$ \Omega(z) = \leftno \big\{ {1\over 4} A (z) \otimes
A(-z) +  {1\over 4} A(-z) \otimes
A (z) - {1\over 8}
-{1\over 4} \leftno \left( T(z)\otimes 1 - 
1 \otimes T(z) \right)^2 \rightno $$
$$- \leftno \left( D_1(z)\otimes 1 - 1\otimes D_1(z) 
\right)
\left( K_1(z)\otimes 1 - 1\otimes K_1(z) \right) \rightno
 \big\}
\times \sum\limits_{r\in\R} K_0(z,r)\otimes K_0(z,-r)
\rightno .$$

 We consider the system of equations
$$ \Omega_k (\tau\otimes\tau) = 0, \hbox{\hskip 1cm}
k\geq -1,$$
on a function $\tau\in F$. In the case of affine
Kac-Moody algebra $A_1^{(1)}$ (i.e., $\dg = sl_2 (\C),
n=0 $), this system is equivalent to the KdV hierarchy
of partial differential equations ([Kac]). 
In the toroidal case considered here, we obtain a 
hierarchy that contains the KdV hierarchy as a proper 
subset.

 The tensor square $F\otimes F$ of a polynomial algebra
$F$ is again a polynomial algebra in twice as many 
variables. Denoting the variables in the first copy of 
$F$ by $q^\p , x_j^\p , u_i^\p , v_i^\p$ and in the
second copy of $F$ by $q^\pp , x_j^\pp , u_i^\pp , 
v_i^\pp$, we obtain the following representation
for $\Omega(z):$
$$ \Omega(z) = \leftno \bigg\{ {1\over 16}
\exp{ 2\sum\limits_{j\in\odd} z^j (x_j^\p - x_j^\pp) }
\exp{ -2\sum\limits_{j\in\odd} {z^{-j} \over j}
 \left( {\d\over \d x_j^\p} - {\d\over \d x_j^\pp}
\right) } $$ 
$$+ {1\over 16}
\exp{ - 2\sum\limits_{j\in\odd} z^j (x_j^\p - x_j^\pp) }
\exp{ 2\sum\limits_{j\in\odd} {z^{-j} \over j}
 \left( {\d\over \d x_j^\p} - {\d\over \d x_j^\pp}
\right) } - {1\over 8} $$
$$ - {1\over 4} \leftno \left( 
\sum\limits_{j\in\odd} j z^j (x_j^\p - x_j^\pp) +
\sum\limits_{j\in\odd} z^{-j} 
\left( {\d\over \d x_j^\p} - {\d\over \d x_j^\pp} \right)
\right)^2 \rightno $$
$$ - \leftno \left( \sum\limits_{i\geq 1} i z^{2i} 
(v_i^\p - v_i^\pp) + \left( q^\p {\d\over\d q^\p} -
q^\pp {\d\over\d q^\pp} \right) +
\sum\limits_{i\geq 1} z^{-2i} \left( {\d\over\d u_i^\p}
- {\d\over\d u_i^\pp} \right) \right) \times$$
$$
\times \left( \sum\limits_{i\geq 1} i z^{2i} (u_i^\p -
u_i^\pp) +
\sum\limits_{i\geq 1} z^{-2i} \left( {\d\over\d v_i^\p}
- {\d\over\d v_i^\pp} \right) \right) \rightno \bigg\} 
\times$$
$$\times \sum\limits_{r\in\R} \left({ q^\p\over q^\pp }
\right)^r \exp{ r\sum\limits_{i\geq 1} 
z^{2i} (u_i^\p - u_i^\pp) }
 \exp{ -r\sum\limits_{i\geq 1} {z^{-2i}\over i}
 \left( {\d\over\d v_i^\p} - {\d\over\d v_i^\pp} \right) }
\rightno .$$

We perform the change of variables:
$$ w = {1\over 2} \ln\left({ q^\p\over q^\pp } \right),
\hskip 0.5cm
\tw = {1\over 2} \ln\left(q^\p q^\pp \right),
\hskip 0.5cm
x_j = {1\over 2} (x_j^\p - x_j^\pp) ,
\hskip 0.5cm
\tx_j = {1\over 2} (x_j^\p + x_j^\pp) ,$$
$$u_i = {1\over 2} (u_i^\p - u_i^\pp) ,
\hskip 0.5cm
\tu_i = {1\over 2} (u_i^\p + u_i^\pp) ,
\hskip 0.5cm
v_i = {1\over 2} (v_i^\p - v_i^\pp) ,
\hskip 0.5cm
\tv_i = {1\over 2} (v_i^\p + v_i^\pp) .$$

Then
$$ {\d\over\d w} =  q^\p {\d\over\d q^\p} -
q^\pp {\d\over\d q^\pp}  , \hskip 1cm
{\d\over\d x_j} = 
 {\d\over \d x_j^\p} - {\d\over \d x_j^\pp} ,$$
$${\d\over\d u_i} = 
 {\d\over\d u_i^\p} - {\d\over\d u_i^\pp} , \hskip 1cm
{\d\over\d v_i} = 
 {\d\over\d v_i^\p} - {\d\over\d v_i^\pp} .$$

The expression $\Omega(z) (\tau\otimes\tau)$  
transforms as follows:
$$ \Omega(z) \tau (q^\p, \x^\p, \u^\p, \v^\p)
\tau (q^\pp, \x^\pp, \u^\pp, \v^\pp) =$$
$$ = \leftno \bigg\{ {1\over 16}
\exp{ 4 \sum\limits_{j\in\odd} z^j x_j }
\exp{ -2\sum\limits_{j\in\odd} {z^{-j} \over j}
  {\d\over \d x_j} } $$ 
$$+ {1\over 16}
\exp{ -4 \sum\limits_{j\in\odd} z^j x_j }
\exp{ 2\sum\limits_{j\in\odd} {z^{-j} \over j}
  {\d\over \d x_j} } - {1\over 8} $$
$$ - {1\over 4} \leftno \left( 
\sum\limits_{j\in\odd} 2j z^j x_j +
\sum\limits_{j\in\odd} z^{-j} {\d\over \d x_j}
\right)^2 \rightno $$
$$ - \leftno \left( \sum\limits_{i\geq 1} 2i z^{2i} v_i +
 {\d\over\d w} +
\sum\limits_{i\geq 1} z^{-2i} {\d\over\d u_i} \right)
\left( \sum\limits_{i\geq 1} 2i z^{2i} u_i +
\sum\limits_{i\geq 1} z^{-2i} {\d\over\d v_i}
 \right) \rightno \bigg\} \times$$
$$\times \sum\limits_{r\in\R} e^{2rw}
 \exp{ 2r\sum\limits_{i\geq 1} 
z^{2i} u_i} \exp{ -r\sum\limits_{i\geq 1} {z^{-2i}\over i}
{\d\over\d v_i} } \rightno $$ 
$$\tau(e^{\tw+w}, \btx + \x, \btu + \u, \btv + \v)
\tau(e^{\tw-w}, \btx - \x, \btu - \u, \btv - \v) .
\eqno{(3.1)}$$

Consider the completion of the group algebra $\C[\R]$:
$\ol{\C[\R]} = \prod\limits_{\r\in\R} \C e^{rw}$.
The element
$$\delta(w) = \sum\limits_{r\in\R} e^{rw}$$
is a formal analog of the $\delta$-function
(cf. [FLM], Sec. 2.2). For any $X(w) \in \C[\R]$ we
have
$$ \delta (w) X(w) = \delta (w) X(0) .\eqno{(3.2)}$$
To establish this identity, it is sufficient to verify
it for the basis elements $X(w) = e^{aw}$:
$$ \delta (w) e^{aw} = \sum\limits_{r\in\R} e^{(r+a)w}
= \delta (w) \cdot 1 .$$

{\bf Proposition 2.} 
{\it
Let $X(w) \in \C[\R]$. Then
}
$$ \left[ \left( {\d\over\d w} \right)^n \delta(w) \right]
X(w) = 
\sum\limits_{k=0}^n (-1)^{n-k} \left( {n \atop k} \right)
\left[ \left( {\d\over\d w} \right)^k \delta(w) \right] 
\left[ \left( {\d\over\d w} \right)^{n-k}
 X(w)\big|_{w=0} \right] .$$

Proof by induction. The basis of induction is (3.2).
To make the inductive step, we use the Leibnitz rule:
$$ \left[ \left( {\d\over\d w} \right)^{n+1} \delta(w) 
\right] X(w) = 
{\d\over\d w} \left( \left[ \left( {\d\over\d w} 
\right)^n \delta(w) \right] X(w) \right) -
\left[ \left( {\d\over\d w} \right)^n \delta(w) \right] 
{\d X \over \d w} =$$
$$ = \sum\limits_{k=0}^n (-1)^{n-k}
 \left( {n \atop k} \right)
\left[ \left( {\d\over\d w} \right)^{k+1} \delta(w) 
\right]
\left[ \left( {\d\over\d w} \right)^{n-k}
 X(w)\big|_{w=0} \right] $$
$$+ \sum\limits_{k=0}^n (-1)^{n-k+1}
 \left( {n \atop k} \right)
\left[ \left( {\d\over\d w} \right)^{k} \delta(w) 
\right]
\left[ \left( {\d\over\d w} \right)^{n-k+1}
 X(w)\big|_{w=0} \right] =$$
$$ = \sum\limits_{k=0}^{n+1} (-1)^{n+1-k} \left( 
{n+1 \atop k} \right)
\left[ \left( {\d\over\d w} \right)^k \delta(w) \right] 
\left[ \left( {\d\over\d w} \right)^{n+1-k}
 X(w)\big|_{w=0} \right] .$$

{\bf Proposition 3.} 
{\it
Let $P(r) = \sum\limits_{n\geq 0}
r^n P_n$, where $P_n \in \Diff (w, y_1, y_2, \ldots)$
are differential operators that may depend on
${\d\over\d w}$ but not on $w$. Suppose that for every
$f (w, \y) = \sum\limits_{i=1}^N e^{r_i w} 
f_i (\y) \in\C[\R] 
\otimes \C[y_1, y_2, \ldots]$ we have $P_n f = 0$
for all but finitely many $n$.
If
$$ \sum\limits_{r\in\R} e^{rw} P(r) g(w, \y) = 0 $$
for some $g(w, \y) \in \C[\R] \otimes 
\C[y_1, y_2, \ldots]$ then
$$ P\left(\e - {\d\over\d w} \right) g(w, \y)
 \bigg|_{w=0} = 0
\hbox{\hskip 1.5cm \it as a polynomial in \ } \e .$$
}

Proof. We have
$$ 0 = \sum\limits_{r\in\R} e^{rw} P(r) g(w,\y) =
\sum\limits_{n\geq 0} \sum\limits_{r\in\R} 
r^n e^{rw} P_n g(w,\y) =$$
$$ = \sum\limits_{n\geq 0} 
\left[ \left( {\d\over\d w} \right)^n \delta(w) \right]
P_n g(w,\y) =$$
$$ = \sum\limits_{n\geq 0} 
\sum\limits_{k=0}^n (-1)^{n-k} \left( {n \atop k} \right)
\left[ \left( {\d\over\d w} \right)^k \delta(w) \right] 
\left[ \left( {\d\over\d w} \right)^{n-k}
P_n g(w,\y)\big|_{w=0} \right] =$$
$$ = \sum\limits_{k\geq 0}
\left[ \left( {\d\over\d w} \right)^k \delta(w) \right] 
\sum\limits_{m = n-k \geq 0} (-1)^m \left( {m+k \atop k}
\right) \left[ \left( {\d\over\d w} \right)^{m}
P_{m+k} g(w,\y)\big|_{w=0} \right] .$$
The last expression is a finite linear combination
of $\left\{ \left( {\d\over\d w}\right)^k \delta(w) \right\}$.
Since these are linearly independent then for every $k$
$$ \sum\limits_{m\geq 0} (-1)^m \left( {m+k \atop k}
\right) \left[ \left( {\d\over\d w} \right)^{m}
P_{m+k} g(w,\y)\big|_{w=0} \right] = 0 .$$
Hence the following polynomial in $\e$ is zero:
$$ 0 = \sum\limits_{k\geq 0} \e^k 
\sum\limits_{m\geq 0} (-1)^m \left( {m+k \atop k}
\right) \left[ \left( {\d\over\d w} \right)^{m}
P_{m+k} g(w,\y)\big|_{w=0} \right] = $$
$$ = P\left(\e - {\d\over\d w} \right) g(w, \y)
\bigg|_{w=0} .$$

We will use this proposition in order to 
derive Hirota bilinear equations from the Casimir
operator equation
$$\Omega_k (\tau \otimes \tau) = 0 .$$ Let us recall
the definition of the Hirota bilinear equations.

For a polynomial $R(y_1, y_2, \ldots)$ and a function
$\tau(\ty_1, \ty_2, \ldots)$ we denote by \break
$R( H_{\ty_1}, H_{\ty_2}, \ldots) \circ  
\tau(\ty_1, \ty_2, \ldots) \tau(\ty_1, \ty_2, \ldots)$
the expression 
$$ R( {\d\over\d y_1} , {\d\over\d y_2} , \ldots)
\tau(\ty_1 + y_1, \ty_2 + y_2, \ldots)
\tau(\ty_1 - y_1, \ty_2 - y_2, \ldots) \big|_{y_i = 0} .$$
Equation
$$R( H_{\ty_1}, H_{\ty_2}, \ldots) \circ  
\tau(\bty) \tau(\bty) = 0$$
is called a Hirota bilinear equation.

Using the same technics as in [Kac], we can transform
(3.1) in the Hirota form with respect to the variables
$\btx, \btu, \btv$. This is based on the following
observation:
$$ R({\d\over\d y_1} , {\d\over\d y_2} , \ldots)
\tau(\bty + \y)
\tau(\bty - \y) = $$
$$ = R({\d\over\d x_1} , {\d\over\d x_2} , \ldots)
\tau(\bty + (\x + \y))
\tau(\bty - (\x + \y))
\bigg|_{\x = 0} =$$
$$ =  R({\d\over\d x_1} , {\d\over\d x_2} , \ldots)
\exp{ \sum\limits_{i\geq 1} y_i {\d\over\d x_i}}
\tau(\bty + \x)\tau(\bty - \x)
\bigg|_{\x = 0} =$$
$$ = R( H_{\ty_1}, H_{\ty_2}, \ldots)
\exp{ \sum\limits_{i\geq 1} y_i H_{\ty_i} } \circ 
\tau(\bty) \tau(\bty) . \eqno{(3.3)}$$

Note that the
operator $\Omega_k = \Omega_k (r)$ in (3.1)
satisfies the conditions of Proposition 3. 
The normal ordering in (3.1) guarantees that all the
differentiations are performed before the multiplications
by $x_j, u_i, v_i$. Thus applying (3.3) and Proposition 3
we obtain that $\Omega (z) \tau\otimes\tau$ can be written
in the Hirota form as follows:
$$  \Omega (z) \tau\otimes\tau =$$
$$ = \bigg\{ {1\over 16}
\exp{ 4 \sum\limits_{j\in\odd} z^j x_j }
\exp{ -2\sum\limits_{j\in\odd} {z^{-j} \over j}
  H_{\tx_j} } $$ 
$$+ {1\over 16}
\exp{ -4 \sum\limits_{j\in\odd} z^j x_j }
\exp{ 2\sum\limits_{j\in\odd} {z^{-j} \over j}
  H_{\tx_j} } - {1\over 8} $$
$$ - {1\over 4} \left( 
\sum\limits_{j\in\odd} 2j z^j x_j +
\sum\limits_{j\in\odd} z^{-j} H_{\tx_j}
\right)^2 $$
$$ - \left( \sum\limits_{i\geq 1} 2i z^{2i} v_i +
 H_{\tw} +
\sum\limits_{i\geq 1} z^{-2i} H_{\tu_i} \right)
\left( \sum\limits_{i\geq 1} 2i z^{2i} u_i +
\sum\limits_{i\geq 1} z^{-2i} H_{\tv_i}
 \right)  \bigg\} \times$$
$$\times  \exp{ \left( \e - H_{\tw} \right)
 \sum\limits_{i\geq 1} z^{2i} u_i} \exp{ -{1\over 2} 
\left( \e - H_{\tw} \right)
\sum\limits_{i\geq 1} {z^{-2i}\over i}
H_{\tv_i} } \times$$
$$\times  \exp{ \sum\limits_{j\in\odd} x_j H_{\tx_j} }
\exp{ \sum\limits_{i\geq 1} u_i H_{\tu_i} }
\exp{ \sum\limits_{i\geq 1} v_i H_{\tv_i} } \circ$$
$$ \tau(e^{\tw}, \btx , \btu , \btv)
\tau(e^{\tw}, \btx , \btu , \btv) .
\eqno{(3.4)}$$

 The solutions for $\Omega_k \tau\otimes\tau = 0$ 
that will be constructed in the next section do not
depend on $\v$, so we can make a reduction 
$H_{\tv_i} = 0$. Also, to simplify the notations we will
denote $H_{\tw}$ by $H_0$, $H_{\tx_{2i+1}}$ by $H_{2i+1}$
and $H_{\tu_i}$ by $H_{2i}$. We get
$$ \sum\limits_{k\in\Z} \Omega_k z^{-2k}
 \tau\otimes\tau =$$
$$= \bigg\{ {1\over 16}
\exp{ 4 \sum\limits_{j\in\odd} z^j x_j }
\exp{ -2\sum\limits_{j\in\odd} {z^{-j} \over j}
  H_{j} } $$ 
$$+ {1\over 16}
\exp{ -4 \sum\limits_{j\in\odd} z^j x_j }
\exp{ 2\sum\limits_{j\in\odd} {z^{-j} \over j}
  H_{j} } - {1\over 8} $$
$$ - {1\over 4} \left( 
\sum\limits_{j\in\odd} 2j z^j x_j +
\sum\limits_{j\in\odd} z^{-j} H_{j}
\right)^2 $$
$$ - \left( \sum\limits_{i\geq 1} 2i z^{2i} v_i +
 H_{0} +
\sum\limits_{i\geq 1} z^{-2i} H_{2i} \right)
\left( \sum\limits_{i\geq 1} 2i z^{2i} u_i
 \right)  \bigg\} \times$$
$$\times  \exp{ \left( \e - H_0\right) 
\sum\limits_{i\geq 1} 
z^{2i} u_i} \exp{ \sum\limits_{j\in\odd} x_j H_j }
\exp{ \sum\limits_{i\geq 1} u_i H_{2i} } \circ$$
$$ \tau(e^{\tw}, \btx , \btu)
\tau(e^{\tw}, \btx , \btu)
\eqno{(3.5)}$$ 

This can be interpreted as a formal series in independent
variables $z^2, \e, x_j, u_i, v_i$ with the coefficients
being Hirota polynomials in $H_0, H_1, H_2, \ldots$. 
We obtain a hierarchy of the Hirota bilinear
equations by considering coefficients at various
monomials in the equations $\Omega_k \circ
\tau(e^{\tw}, \btx , \btu) \tau(e^{\tw}, \btx , \btu)
= 0, \ \ k\geq -1$. We call this system of Hirota 
equations the extended KdV hierarchy.

 When we consider Hirota equations corresponding
to the monomials that depend on $x_1, x_3, \ldots$
only, we recover the KdV hierarchy. Other non-trivial
equations of degrees less or equal to 5 are given
below.

 From the coefficient at $x_1 u_2$:
$$ H_0 H_1^3 + 2 H_0 H_3 - 6 H_1 H_2 = 0 . \eqno{(3.6)}$$ 
From the coefficient at $u_2^2$:
$$ H_0^2 H_1^4 - 4 H_0^2 H_1 H_3 + 48 H_0 H_4 - 48 H_2^2
 = 0 . \eqno{(3.7)}$$ 
From the coefficient at $x_1 u_4 $:
$$H_0 H_1^5 + 20 H_0 H_1^2 H_3 + 24 H_0 H_5 - 120 H_1 H_4
 = 0 . \eqno{(3.8)}$$ 
From the coefficient at $\e x_1 u_2^2$:
$$ H_0 H_1^5 + 20 H_0 H_1^2 H_3 + 24 H_0 H_5 -
{40 \over 3} H_1^3 H_2 - {80 \over 3} H_2 H_3 - 40 H_1 H_4
 = 0 . \eqno{(3.9)}$$ 
From the coefficient at $u_2 x_3$:
$$  H_0 H_1^5 + 5 H_0 H_1^2 H_3 + 24 H_0 H_5 -
5 H_1^3 H_2 - 40 H_2 H_3 
 = 0 . \eqno{(3.10)}$$

The equations (3.8)-(3.10) form a basis in the space
of equations of degree 5 in this hierarchy. The following
Hirota equations belong to this space:
$$ H_1^3 H_2 - H_0 H_1^2 H_3 = 0 , \eqno{(3.11)}$$
$$ H_1^3 H_2 + 2 H_2 H_3 - 6 H_1 H_4 = 0 . \eqno{(3.12)}$$

Equations (3.6) and (3.12) coincide up to a
relabeling of the variables. An equation, equivalent to (3.12)
occurs as the second equation (of degree 5) in the
Kadomtsev-Petviashvili
(KP) hierarchy. The equation (3.11) is apparently new. 

\

{\bf 4. $N$-soliton solutions.}

The same method as in [Kac] allows us to
construct solutions for the extended KdV hierarchy.
The idea of this method is based on the following

{\bf Proposition 4.} 
{\it
The operator $\big(1 + \lambda A (z,r)
\big) \otimes \big(1 + \lambda A (z,r) \big)$ commutes 
with $\Omega_k$.
}

Proof. Using Proposition 3.4.2 in [FLM] we obtain

$$ A (z_1, r_1) A (z_2, r_2) = 
A (z_1) A (z_2) K_0 (z_1, r_1) K_0 (z_2, r_2) =$$
$$ = \left( {z_1 - z_2 \over z_1 + z_2} \right)^2 
\leftno A (z_1) A(z_2) \rightno
K_0 (z_1, r_1) K_0 (z_2, r_2) .\eqno{(4.1)}$$

By Proposition 1, $\Omega_k$ commutes with 
$$A (z,r) \otimes 1 + 1 \otimes A (z,r) $$
and thus with 
$$\left(
A (z_1,r_1) \otimes 1 + 1 \otimes A (z_1,r_1) \right)
\left( A (z_2,r_2) \otimes 1 + 1 \otimes A (z_2,r_2) 
\right).$$

Passing to the limit with $z_1 \to z, z_2 \to z$, 
taking (4.1) into account and setting
$r_1 = r_2 = r$, we get that
$$ A(z,r) \otimes A(z,r)$$
commutes with $\Omega_k$. But then
$$\big(1 + \lambda A (z,r)
\big) \otimes \big(1 + \lambda A (z,r) \big) =$$
$$ = 1 \otimes 1 + \lambda \left(A (z,r) 
\otimes 1 + 1 \otimes A (z,r) \right)
+ \lambda^2 A(z,r) \otimes A(z,r)$$
commutes with $\Omega_k$ and the proposition is proved.

{\bf Corollary 5. } 
{\it
If $\tau$ is a solution of $\Omega_k (\tau
\otimes \tau) = 0$ then $(1 + \lambda A(z,r) )\tau$ is 
also a solution of this equation.
}

{ Proof.} Indeed,
$$0 = \left( \big( 1 + \lambda A (z,r) \big) \otimes 
\big( 1 + \lambda A (z,r) \big)\right)  
\Omega_k (\tau \otimes \tau) = \Omega_k 
\left( \left(1 + \lambda A (z,r) ) \tau \otimes
(1 + \lambda A (z,r) \right) \tau \right) .$$

{\bf Lemma 6.} 
{\it The function
$\tau = 1$ is a solution of $\Omega_k (\tau \otimes \tau)$
for $k \geq -1$.
}

{Proof.}
Recalling (3.1) we obtain 
$$\Omega(z) (1\otimes 1) = \bigg\{
{1\over 16} 
\exp{ 4 \sum\limits_{j\in\odd} z^j x_j }
   +{1\over 16}
\exp{ -4 \sum\limits_{j\in\odd} z^j x_j }
 - {1\over 8}$$
$$ -  {1\over 4} \left( 
\sum\limits_{j\in\odd} 2j z^j x_j \right)^2 -
\left( \sum\limits_{i\geq 1} 2i z^{2i} v_i \right)
\left( \sum\limits_{i\geq 1} 2i z^{2i} u_i \right)
 \bigg\}  \times$$
$$ \times \sum\limits_{r\in\R} e^{rw}
 \exp{ r\sum\limits_{i\geq 1} 
z^{2i} u_i} .$$
We can see that only non-negative powers of $z$ appear
in the right hand side. Moreover,
the coefficients at $z^0$ and $z^2$ vanish,
thus $\tau = 1$ is a solution of $\Omega_k (\tau\otimes\tau) = 0$
for $k\geq -1$.

{\bf Remark.} $\Omega_k (1 \otimes 1)$ for $k<-1$
has an interesting representation-theoretic
meaning - it gives a vacuum vector inside 
$\ol{F\otimes F}$.

{\bf Theorem 7.}
{\it
 For $\lambda_1\ldots \lambda_N, z_1,\ldots,z_N, r_1,\ldots,r_N \in\R$
the function
$$ \tau (w, x_1, x_3, x_5, \ldots, u_1, u_2, \ldots ) =$$
$$= \sum\limits_{{ 0 \leq k \leq N}\atop{1\leq i_1 < \ldots < i_k \leq N}}
\lambda_{i_1} \ldots \lambda_{i_k} 
\prod\limits_{1\leq\mu < \nu\leq k}
\left({ z_{i_\mu} - z_{i_\nu} \over z_{i_\mu} + z_{i_\nu} 
} \right)^2 \times$$
$$\times
\exp{ \sum\limits_{m = 1}^k r_{i_m} w
+ 2 \sum\limits_{j\in\odd} \sum\limits_{m = 1}^k z_{i_m}^j x_j
+ \sum\limits_{j\in\N} \sum\limits_{m = 1}^k 
r_{i_m} z_{i_m}^{2j} u_j } $$
is a solution of the extended KdV hierarchy.
}

{Proof.}
Using Lemma 6 and Corollary 5, we obtain that 
$$ \tau = ( 1 + 2 \lambda_1 A(z_1, r_1) ) \ldots 
  ( 1 + 2 \lambda_N A(z_N, r_N) ) 1  = $$
$$ = \sum\limits_{{ 0 \leq k \leq N}\atop{1\leq i_1 < \ldots < i_k \leq N}}
\lambda_{i_1} \ldots \lambda_{i_k} 
 2^k A(z_{i_1}, r_{i_1}) \ldots A(z_{i_k}, r_{i_k}) 1
\eqno{(4.2)}$$
is a solution of the extended KdV hierarchy.
Using Proposition 3.4.1 of [FLM], we get a generalization
of (4.1):
$$  A(z_{i_1}, r_{i_1}) \ldots A(z_{i_k}, r_{i_k}) = 
\prod\limits_{1\leq\mu < \nu\leq k}
\left({ z_{i_\mu} - z_{i_\nu} \over z_{i_\mu} + z_{i_\nu} 
} \right)^2
\leftno A(z_{i_1}, r_{i_1}) \ldots A(z_{i_k}, r_{i_k}) 
\rightno . \eqno{(4.3)}$$
Recalling that
$$A(z,r) = $$
$$ = {1\over 2} \exp{rw + 2\sum\limits_{j\in\odd} z^j x_j + r\sum\limits_{i\in\N} z^{2i} u_i}
\exp{-2\sum\limits_{j\in\odd} {z^{-j}\over j} 
{\d\over\d x_j }
 -r\sum\limits_{i\in\N} {z^{-2i}\over i} 
{\d\over\d v_i} }$$
and using the definition of the normal ordering, we get
$$ 2^k \leftno
 A(z_{i_1}, r_{i_1}) \ldots A(z_{i_k}, r_{i_k}) 
\rightno 1 =$$
$$= \exp{ \sum\limits_{m = 1}^k r_{i_m} w
+ 2 \sum\limits_{j\in\odd} \sum\limits_{m = 1}^k z_{i_m}^j x_j
+ \sum\limits_{j\in\N} \sum\limits_{m = 1}^k 
r_{i_m} z_{i_m}^{2j} u_j }. \eqno{(4.4)}$$
Combining (4.2), (4.3) and (4.4), we obtain the claim of the theorem.

Let us finish this section by presenting the formulas
for the solutions of the partial differential equations corresponding
to Hirota equations (3.6) and (3.11). Setting $x=x_1, y = w, z = x_3,
t = u_1$, we rewrite (3.6) and (3.11) as
$$  H_x^3 H_y + 2 H_y H_z - 6 H_x H_t = 0 , \eqno{(4.5)} $$
$$ H_x^3 H_t - H_x^2 H_y H_z = 0 . \eqno{(4.6)} $$

Every bilinear Hirota equation can be written as a partial differential
equation via the logarithmic transformation.
Introducing $f(t,x,y,z) = {\d\over\d x} \ln \tau$, we can write
(4.5) as a non-linear PDE:
$$ {\d\over\d x}( 6 f_t - f_{xxy} - 6 f_x f_y ) - f_{yz} = 0 .\eqno{(4.7)}$$
Applying the previous theorem and treating all non-effective
variables as parameters, we obtain the following solutions of (4.7):
$$ f(t,x,y,z) = {\d\over\d x} \ln \tau , \eqno{(4.8)}$$
where 
$$\tau = 
\sum\limits_{{ 0 \leq k \leq N}\atop{1\leq i_1 < \ldots < i_k \leq N}}
\lambda_{i_1} \ldots \lambda_{i_k} 
\prod\limits_{1\leq\mu < \nu\leq k}
\left({ c_{i_\mu} - c_{i_\nu} \over c_{i_\mu} + c_{i_\nu} 
} \right)^2 \times$$
$$\times
\exp{ 
\left(\sum\limits_{m = 1}^k  r_{i_m} c_{i_m}^{2}\right) t 
+ 2 \left(\sum\limits_{m = 1}^k c_{i_m}\right) x + 
\left(\sum\limits_{m = 1}^k r_{i_m}\right) y
+ 2 \left(\sum\limits_{m = 1}^k c_{i_m}^3\right) z  }.
\eqno{(4.9)}$$

The second derivative of \ \ $\ln \tau$ \ \ exibits the $N$-soliton behaviour,
thus (4.8-4.9) is the $N$-soliton potential.

Note that these solutions are different from the solutions
arising in the context of the KP hierarchy. This may indicate that there
is a larger Lie algebra governing the symmetries of this equation. 

The transformation \ \ $\tau = e^g$, \  $g = \ln \tau $ 
\ \ allows us to write Hirota equation (4.6) as a PDE:

$$ g_{xxxt} + 6 g_{xx} g_{xt} - g_{xxyz} - 4 g_{xy} g_{xz} - 2 g_{xx} g_{yz} = 0 . \eqno{(4.10)}$$

The $\tau$-function expression (4.9) provides a family of solutions
of (4.10).

\

{\bf 5. Appendix. Generalized Casimir operators.}

The Casimir operator plays a prominent role in the 
representation theory of Lie algebras. The main feature
of the Casimir operator is that it commutes with the 
action of the Lie algebra. In order to construct it 
one needs a non-degenerate invariant bilinear form
on the Lie algebra (see Section 2.8 in [Kac]).
The semidirect product of $\g$ with ${\cal D}$ possesses
such a form, however we deal here with a deformation of
this algebra. Nevertheless, the corresponding operator
still commutes with the action of $\g$ (but not
${\cal D}$).

First, we introduce a Casimir operator for $\dg$.
Let $\{ e^i \}$ and $\{ f^i \}$ be dual bases in $\dg$
with respect to the invariant bilinear form.
Then for every $X\in\dg$
$$ \sum\limits_i ( X | e^i ) f^i =
\sum\limits_i ( X | f^i ) e^i = X .$$
Since $( \dg_{j_1} | \dg_{j_2} ) = 0$ unless
$j_2 = -j_1 (\mod h)$ then for $X\in\dg_m$ we have
$$ \sum\limits_i ( X | e^i_{-m} ) f^i_m =
\sum\limits_i ( X | f^i_{-m} ) e^i_m = X . \eqno{(5.1)}$$

The Casimir element $\dot\Omega$ is defined by
$$ \dot\Omega =  \sum\limits_i
e^i\otimes f^i = \sum\limits_i
\sum\limits_{j_1,j_2\in\Z_h} e^i_{j_1}\otimes f^i_{j_2} 
\in U(\dg)\otimes U(\dg).$$

The algebra $U(\dg)\otimes U(\dg)$ is graded by the 
root lattice of $\dg$. The projection of the root lattice
on $\Z$ by the height function induces a $\Z$-grading
of $U(\dg)\otimes U(\dg)$. Since $\dot\Omega$ commutes
with the action of $\dot\h$, it belongs to the height $0$
component of $U(\dg)\otimes U(\dg)$. Hence the same is
true for the principal $\Z_h$-grading of 
$U(\dg)\otimes U(\dg)$, which means that
$$ \sum\limits_i e^i_{j_1} \otimes f^i_{j_2} = 0
\hbox{\hskip 1cm \rm if \hskip 0.5cm} j_1 + j_2 \neq 0
(\mod h). \eqno{(5.2)}$$
Consequently,
$$\dot\Omega = \sum\limits_i \sum\limits_{j\in\Z_h}
e^i_{j}\otimes f^i_{-j} .$$
The Lie algebra $\dg$ is embedded in 
$U(\dg)\otimes U(\dg)$ via the diagonal map:
$$ X \mapsto X \otimes 1 + 1 \otimes X .$$
The Casimir element $\dot\Omega$ commutes with 
$\dg$:
$$ [ X , \dot\Omega ] = 
\sum\limits_i \sum\limits_{j\in\Z_h} 
\left( [ X , e^i_{j} ] \otimes f^i_{-j} +
e^i_{j} \otimes [ X , f^i_{-j} ] \right) = 0 .$$
In case when $X \in \dg_m$, taking the 
$U(\dg)_{j+m}\otimes U(\dg)_{-j}$-component of the
previous equality, we obtain
$$\sum\limits_i \left( [ X , e^i_{j} ] \otimes f^i_{-j} +
e^i_{j+m} \otimes [ X , f^i_{-j-m} ] \right) = 0 .
 \eqno{(5.3)}$$

Now consider the generalized Casimir operators 
$\Omega_k$, $k\in\Z$, for the toroidal algebra $\g$:
$$\Omega_k = \sum\limits_{\r\in\R^n}
\sum\limits_{j\in\Z} \sum\limits_i
\phi\left(e^i_{j} s^j \t^\r \otimes 
f^i_{-j} s^{kh-j} \t^{-\r}\right)$$
$$ + {1 \over h}  
\sum\limits_{\r\in\R^n} \sum\limits_{j\in\Z}
s^{jh} \t^\r D_s \otimes 
\phi( s^{(k-j)h} \t^{-\r} K_0)$$
$$ + {1 \over h}  
 \sum\limits_{\r\in\R^n} \sum\limits_{j\in\Z}
\phi( s^{jh} \t^\r K_0) \otimes 
s^{(k-j)h} \t^{-\r} D_s$$
$$ + \sum\limits_{\r\in\R^n} 
\sum\limits_{p=1}^n \sum\limits_{j\in\Z}
s^{jh} \t^\r D_p \otimes 
\phi( s^{(k-j)h} \t^{-\r} K_p)$$
$$ + \sum\limits_{\r\in\R^n} 
\sum\limits_{p=1}^n \sum\limits_{j\in\Z}
\phi( s^{jh} \t^\r K_p) \otimes 
s^{(k-j)h} \t^{-\r} D_p$$ 
$$ - { (\rho | \rho) \over h^2 } 
\sum\limits_{\r\in\R^n} \sum\limits_{j\in\Z}
\phi\left( s^{jh} \t^\r K_0 \otimes
 s^{(k-j)h} \t^{-\r} K_0\right) .$$

\

{\bf Remark.} The constant ${ (\rho | \rho) \over h^2 }$
is chosen to make $\tau = 1$ a solution of
$\Omega_0 (\tau \otimes \tau) = 0$. 
Using the Freudenthal - de Vries ``strange''
formula ([Kac], (12.1.8)) we obtain that
$${ (\rho | \rho) \over h^2 } = {\dim\dg \over 12 h}
= { \ell ( h + 1) \over 12 h } .$$

{\bf Proposition 1.} 
{\it
The operators 
$$\Omega_k : \hbox{\hskip 0.5cm}
 F\otimes F \rightarrow \overline{F\otimes F}$$
commute with the action of $\g$.
}

{Proof.} Since $\g$ is generated by the elements
$X s^{m_0} \t^\m$ with $X\in\dg_{m_0}$ then it is 
sufficient to show that 
$[ \phi( X s^{m_0} \t^\m) , \Omega_k ] = 0$.
Indeed,
$$  [ \phi( X s^{m_0} \t^\m) , \Omega_k ] =
 \sum\limits_{\r\in\R^n}
\sum\limits_{j\in\Z} \sum\limits_i
\phi\left( [  X s^{m_0} \t^\m , e^i_{j} s^j \t^\r ]
\otimes f^i_{-j} s^{kh-j} \t^{-\r} \right)$$
$$ + \sum\limits_{\r\in\R^n}
\sum\limits_{j\in\Z} \sum\limits_i
\phi\left( e^i_{j} s^j \t^\r \otimes 
[ X s^{m_0} \t^\m , f^i_{-j} s^{kh-j} \t^{-\r} ]\right)$$
$$ - {1 \over h}  
\sum\limits_{\r\in\R^n} \sum\limits_{j\in\Z}
[ s^{jh} \t^\r D_s ,  \phi( X s^{m_0} \t^\m) ] \otimes 
\phi( s^{(k-j)h} \t^{-\r} K_0)$$
$$ - {1 \over h} 
 \sum\limits_{\r\in\R^n} \sum\limits_{j\in\Z}
\phi( s^{jh} \t^\r K_0) \otimes 
[ s^{(k-j)h} \t^{-\r} D_s , \phi( X s^{m_0} \t^\m) ]$$
$$ - \sum\limits_{\r\in\R^n} 
\sum\limits_{p=1}^n \sum\limits_{j\in\Z}
[ s^{jh} \t^\r D_p , \phi( X s^{m_0} \t^\m) ] \otimes 
\phi( s^{(k-j)h} \t^{-\r} K_p)$$
$$ - \sum\limits_{\r\in\R^n} 
\sum\limits_{p=1}^n \sum\limits_{j\in\Z}
\phi( s^{jh} \t^\r K_p) \otimes 
[s^{(k-j)h} \t^{-\r} D_p , \phi( X s^{m_0} \t^\m) ] = $$ 
$$= \sum\limits_{\r\in\R^n} \sum\limits_{j\in\Z}
\sum\limits_i \phi\left( 
[ X , e^i_{j}] s^{j+m_0} \t^{\r+\m} \otimes 
f^i_{-j} s^{kh-j} \t^{-\r}\right)$$
$$ + \sum\limits_{\r\in\R^n} \sum\limits_{j\in\Z}
\sum\limits_i \phi\left(
e^i_{j+m_0} s^{j+m_0} \t^{\r+\m} \otimes
[X, f^i_{-j-m_0} ] s^{kh-j} \t^{-\r} \right)$$
$$ + \sum\limits_{\r\in\R^n} \sum\limits_{j\in\Z}
\phi \left( \sum\limits_i ( X | e^i_{-m_0} ) 
\left\{ {m_0 \over h} s^{jh} \t^{\r+\m} K_0 +
\sum\limits_{p=1}^n m_p s^{jh} \t^{\r+\m} K_p \right\}
\otimes 
f^i_{m_0} s^{(k-j)h+m_0} \t^{-\r} \right)$$
$$ + \sum\limits_{\r\in\R^n} \sum\limits_{j\in\Z}
\phi \left(
\sum\limits_i e^i_{m_0} s^{jh+m_0} \t^\r \otimes
( X | f^i_{-m_0} ) 
\left\{ {m_0 \over h} s^{(k-j)h} \t^{\m-\r} K_0 +
\sum\limits_{p=1}^n m_p s^{(k-j)h} \t^{\m-\r} K_p \right\}
\right)$$
$$ - {m_0 \over h}  
\sum\limits_{\r\in\R^n} \sum\limits_{j\in\Z}
\phi\left( X s^{jh+m_0} \t^{\r+\m}  \otimes 
 s^{(k-j)h} \t^{-\r} K_0\right)$$
$$ - {m_0 \over h} 
 \sum\limits_{\r\in\R^n} \sum\limits_{j\in\Z}
\phi\left( s^{jh} \t^\r K_0 \otimes 
 X s^{(k-j)h+m_0} \t^{\m-\r} \right)$$
$$ - \sum\limits_{\r\in\R^n} 
\sum\limits_{p=1}^n \sum\limits_{j\in\Z}
m_p \phi\left( X s^{jh+m_0} \t^{\r+\m} \otimes 
s^{(k-j)h} \t^{-\r} K_p \right)$$
$$ - \sum\limits_{\r\in\R^n} 
\sum\limits_{p=1}^n \sum\limits_{j\in\Z}
m_p \phi\left( s^{jh} \t^\r K_p \otimes 
X s^{(k-j)h+m_0} \t^{\m-\r} \right) . $$
Applying (5.3) we see immediately that
the sum of the first two terms is zero while the rest
cancel out due to (5.1).

In order to obtain explicit expression (2.14) for
$\Omega(z)$, we use the following dual bases in $\dg$
(see Theorem 2.2 in [Kac] and (1.1)):
$$\left\{ T_i , A^\alpha  
\right\}^{i=1,\ldots,\ell}_{\alpha\in\dot\Delta_s}
\hbox{\it \hskip 1cm and \hskip 1cm }
\left\{  {1\over h} T_{\ell + 1 - i} , 
{1\over (A^\alpha | A^{-\alpha} )}
 A^{-\alpha} 
\right\}^{i=1,\ldots,\ell}_{\alpha\in\dot\Delta_s} .$$

 Using these bases 
% and formulas (2.1)-(2.13), 
we obtain the following generating series for the family 
of generalized Casimir operators:

$$\Omega(z) = \sum\limits_{k\in\Z} \Omega_k z^{-kh} =$$
$$ = {1\over h} \sum\limits_{i=1}^\ell 
\sum\limits_{\r\in\R^n} \sum\limits_{j\in\Z}
\sum\limits_{k\in\Z}
\phi\left( T_i s^{m_i + jh} \t^\r \otimes
T_{\ell + 1 - i} s^{-m_i + (k-j)h} \t^{-\r} \right) 
z^{-kh} $$
$$ + \sum\limits_{\alpha\in\dot\Delta_s} 
{1 \over ( A^\alpha | A^{-\alpha} ) }
\sum\limits_{\r\in\R^n} \sum\limits_{j\in\Z}
\sum\limits_{k\in\Z}
\phi\left( A^\alpha_j s^j \t^\r \otimes
A^{-\alpha}_{-j} s^{-j + kh} \t^{-\r} \right) z^{-kh} $$
$$ + {1 \over h}  
\sum\limits_{\r\in\R^n} \sum\limits_{j\in\Z}
\sum\limits_{k\in\Z}
s^{jh} \t^\r D_s \otimes 
\phi( s^{(k-j)h} \t^{-\r} K_0) z^{-kh} $$
$$ + {1 \over h}  
 \sum\limits_{\r\in\R^n} \sum\limits_{j\in\Z}
\sum\limits_{k\in\Z}
\phi( s^{jh} \t^\r K_0) \otimes 
s^{(k-j)h} \t^{-\r} D_s z^{-kh} $$
$$ + \sum\limits_{\r\in\R^n} 
\sum\limits_{p=1}^n \sum\limits_{j\in\Z}
\sum\limits_{k\in\Z}
s^{jh} \t^\r D_p \otimes 
\phi( s^{(k-j)h} \t^{-\r} K_p) z^{-kh} $$
$$ + \sum\limits_{\r\in\R^n} 
\sum\limits_{p=1}^n \sum\limits_{j\in\Z}
\sum\limits_{k\in\Z}
\phi( s^{jh} \t^\r K_p) \otimes 
s^{(k-j)h} \t^{-\r} D_p z^{-kh} $$ 
$$ - { \ell (h + 1) \over 12 h } 
\sum\limits_{\r\in\R^n} \sum\limits_{j\in\Z}
\sum\limits_{k\in\Z}
\phi\left( s^{jh} \t^\r K_0 \otimes
 s^{(k-j)h} \t^{-\r} K_0\right)  z^{-kh} .$$

Taking (5.2) into account, we get
$$ \sum\limits_{j\in\Z} \sum\limits_{k\in\Z}
\phi\left( A^\alpha_j s^j \t^\r \otimes
A^{-\alpha}_{-j} s^{-j + kh} \t^{-\r} \right) z^{-kh} =
\sum\limits_{j_1\in\Z} \sum\limits_{j_2\in\Z}
\phi\left( A^\alpha_{j_1} s^{j_1} \t^\r \otimes
A^{-\alpha}_{j_2} s^{j_2} \t^{-\r} \right) 
z^{j_1 + j_2} .$$
Thus
$$\Omega(z) =  \sum\limits_{\r\in\R^n} \bigg(
 {1\over h} \sum\limits_{i=1}^\ell 
T_i (z, \r) \otimes T_{\ell + 1 - i} (z, -\r)
+ \sum\limits_{\alpha\in\dot\Delta_s} 
{1 \over ( A^\alpha | A^{-\alpha} ) }
A^\alpha (z, \r) \otimes A^{-\alpha}  (z, -\r)$$
$$ + {1\over h}  D_s (z, \r) \otimes K_0 (z, -\r)
 + {1\over h} K_0(z, \r) \otimes D_s (z, -\r) $$
$$+ \sum\limits_{p=1}^n D_p (z, \r) \otimes K_p (z, -\r)
+  \sum\limits_{p=1}^n K_p (z, \r) \otimes D_p (z, -\r) 
- {\ell ( h + 1 ) \over 12 h }  K_0(z, \r)
\otimes K_0 (z, -\r) \bigg) =$$
$$ = \leftno \bigg\{ {1\over h} 
\sum\limits_{i=1}^\ell T_i(z)\otimes T_{\ell+1-i}(z)
+ \sum\limits_{\alpha\in\dot\Delta_s} 
{1 \over ( A^\alpha | A^{-\alpha} ) }
A^{\alpha} (z) \otimes A^{-\alpha} (z) 
- {\ell ( h + 1 ) \over 12 h } $$
$$ + {1\over h} D_s (z) \otimes 1 + {1\over h} 1 \otimes
D_s (z) + \sum\limits_{p=1}^n D_p (z) \otimes K_p (z)
+  \sum\limits_{p=1}^n K_p (z) \otimes D_p (z) \bigg\}
\times $$
$$ \times \sum\limits_{\r\in\R^n} K_0(z,\r) \otimes 
K_0(z,-\r) \rightno .$$

\vfill\eject

{\bf References}

\

\item{[B]} Y.~Billig, {\it Principal vertex operator 
representations for toroidal Lie algebras,} 
preprint 
\break hep-th/9703002.

\item{[DJKM]} E.~Date, M.~Jimbo, M.~Kashiwara, T.~Miwa,
{\it Operator approach to the Kadomtsev~-~Petviashvili
equation. Transformation groups for 
soliton equations III,} J. Phys. Soc. Japan, {\bf 50}
(1981), 3806-3812.

\item{[DS]} V.G.~Drinfeld, V.V.~Sokolov, {\it
Equations of the Koteweg - de Vries type and simple
Lie algebras,} Doklady AN SSSR, {\bf 258} (1981), 11-16.

\item{[EM]} S.~Eswara Rao, R.V.~Moody, {\it Vertex representations for $n$-toroidal
Lie algebras and a generalization of the Virasoro algebra,} 
Comm. Math. Phys. {\bf 159} (1994), 239-264.

\item{[FLM]} I.~Frenkel, J.~Lepowsky, A.~Meurman, {\it Vertex operator 
algebras and the Monster,} Academic Press, Boston, 1989.

% \item{[IKUX]} T.~Inami, H.~Kanno, T.~Ueno, C.-S.~Xiong,
% {\it Two-toroidal Lie algebra as current algebra
% of four-dimensional K\"ahler WZW model,}
% hepth/9610187. 

\item{[Kac]} V.G.~Kac, {\it Infinite-dimensional Lie algebras,} 3rd ed.,
Cambridge University Press, Cambridge 1990.

\item{[KW]} V.G.~Kac, M.~Wakimoto, {\it Exceptional 
hierarchies of soliton equations,} Proc. Symposia in
 Pure Math., {\bf 49} (1989), 191-237.

\item{[Kas]} C.~Kassel, {\it K\"ahler differentials and coverings of complex
simple Lie algebras extended over a commutative algebra,}
J. Pure Appl. Algebra {\bf 34} (1985), 265-275.

\item{[Kos]} B.~Kostant, {\it The principal 
three-dimensional subgroup and the Betti numbers of 
a complex simple Lie group,} Amer. J. Math. {\bf 81}
(1959), 973-1032.

\item{[MEY]} R.V.~Moody, S.~Eswara Rao, T.~Yokonuma, {\it Toroidal Lie algebras 
and vertex representations,} Geom. Ded., {\bf 35} (1990), 283-307.

\item{[S]} M.~Sato, {\it Soliton equations as dynamical
systems on infinite-dimensional Grassmann manifolds,}
RIMS Kokyuroku, {\bf 439} (1981), 30-46.

\end